\documentclass[a4paper,11pt]{article}
\usepackage{pos}
\usepackage[capitalise]{cleveref}
\usepackage{subcaption}

\crefname{section}{Sec.}{Secs.}
\crefname{table}{Tab.}{Tabs.}
\crefname{figure}{Fig.}{Figs.}
\crefname{equation}{Eq.}{Eqs.}
\crefname{appendix}{Appendix}{Appendix}

\newcommand{\rescol}[2]{\textcolor{blue}{#1} \textcolor{red}{#2}}

\newcommand{\SO}{\text{SO}}
\newcommand{\SU}{\text{SU}}
\newcommand{\U}{\text{U}}
\newcommand{\Sp}{\text{Sp}}

\def\gsim{\raise0.3ex\hbox{$\;>$\kern-0.75em\raise-1.1ex\hbox{$\sim\;$}}}
\def\lsim{\raise0.3ex\hbox{$\;<$\kern-0.75em\raise-1.1ex\hbox{$\sim\;$}}}

\title{Hunting scalar partners of the Higgs boson at the LHC}

\author*{Werner Porod}

\affiliation{Institut f\"ur Theoretische Physik und Astrophysik, Uni W\"urzburg,
Emil-Hilb-Weg 22,
D-97074 W\"urzburg, Germany}

\emailAdd{porod@physik.uni-wuerzburg.de}

\abstract{Composite Higgs models with a fermionic ultraviolet completion predict in general additional pseudo Nambu Goldstone bosons beside the Higgs multiplet. In this  contribution we discuss their LHC signatures and present first bounds in simplified models which can also be applied to generic models like multi-Higgs models. We then
demonstrate how these can be combined taking a concrete model based on the SU(5)/SO(5) coset as an example. We use this to show how a proper combination of different channels can lead to an improved bound compared to a single channel analysis.
}

\FullConference{%
  Corfu Summer Institute 2022 "School and Workshops on Elementary Particle Physics and Gravity",\\
  28 August - 1 October, 2022\\
  Corfu, Greece}

\begin{document}
\maketitle

\section{Introduction}

The Standard Model (SM) of particle physics features a single scalar field
being a doublet of weak isospin $\SU(2)_L$ that is responsible for the breaking of 
the electroweak (EW) symmetry \cite{Englert:1964et,Higgs:1964pj}. 
Once it acquires a vacuum expectation value (vev), a massive physical scalar particle arises, 
the Higgs boson \cite{Higgs:1964pj} discovered in 2012 at the Large Hadron Collider (LHC) experiments \cite{ATLAS:2012yve,CMS:2012qbp}. 
In contrast, most models of new physics contain extended Higgs sectors: for instance, 
minimal supersymmetric models~\cite{Martin:1997ns} and two Higgs doublet models~\cite{Branco:2011iw} feature a second doublet and
the type-II seesaw models~\cite{Schechter:1980gr,Hirsch:2008gh} a zero hypercharge triplet. Scalar triplets appear 
also in the Georgi-Machacek model~\cite{Georgi:1985nv}. 
In all these scenarios, the scalar fields acquire sizeable couplings to the SM gauge bosons 
and fermions via vevs and/or via mixing with the SM Higgs boson. 
Consequently, they are dominantly singly-produced at colliders, 
and most current searches focus on these channels.

Single production of a scalar is always model dependent and it can be suppressed by tuning the 
single-scalar couplings. By contrast, pair production only depends on the gauge quantum 
numbers of the scalars and cannot be tuned to be small. 
The couplings of two scalars from $\SU(2)_L \times \U(1)_Y$ multiplets to the EW gauge bosons stem from 
the covariant derivatives in the scalar kinetic terms and are always present. 
Their presence imply dominant pair production channels via Drell-Yan processes, where two initial state 
quarks merge via an s-channel gauge boson. There is some model dependence if their is mixing in
the scalar sector but this mixing cannot reduce Drell-Yan pair production cross sections of all scalars at 
the same time and some channels are guaranteed to remain sizeable. 

In this contribution, we focus on models where pair production is the dominant  mode for scalars charged 
under $\SU(2)_L \times \U(1)_Y$. Such scenarios appear naturally in composite Higgs models, where 
the Higgs boson is accompanied by additional light states, protected by parities internal to the strong 
sector~\cite{Ferretti:2016upr}. The Higgs boson emerges in these models as a pseudo-Nambu-Goldstone 
boson (pNGB) \cite{Kaplan:1983fs} following the dynamical breaking of the EW symmetry triggered by
 misalignment in a condensing strong dynamics at the TeV scale \cite{Weinberg:1975gm,Dimopoulos:1979es}. 
A minimal model $\SO(5)/\SO(4)$ based only on the global symmetries with exactly four pNGBs matching 
the Higgs doublet components can be constructed \cite{Agashe:2004rs}. However, it is not easy to 
obtain this symmetry pattern in an underlying gauge/fermion theory {\it \`a la} QCD. 
A fermion condensate $\langle \psi \psi \rangle$ can only generate the following 
patterns \cite{Witten:1983tw,Kosower:1984aw}: 
$\SU(2N)/\Sp(2N)$, $\SU(N)/\SO(N)$ or $\SU(N)^2/\SU(N)$ depending on whether the representation of 
$\psi$ under the confining gauge symmetry is pseudo-real, real or complex, respectively. 
The minimal model with custodial symmetry \cite{Georgi:1984af,Agashe:2006at} features 
$\SU(4)/\Sp(4)$ \cite{Ryttov:2008xe,Galloway:2010bp}, and has one additional gauge singlet 
pNGB besides the Higgs doublet. The next-to-minimal cases contain significantly more pNGBs: 
$14$ for $\SU(5)/\SO(5)$ \cite{Dugan:1984hq,Arkani-Hamed:2002ikv} and $15$ for $\SU(4)^2/\SU(4)$ 
\cite{Ma:2015gra,Vecchi:2015fma}. The departure from minimality does not contradict the null results 
of direct searches for physics beyond the SM (BSM) at colliders. The pNGBs are typically 
heavier than the Higgs boson and have only EW interactions, hence being difficult to discover 
at hadron colliders and are too heavy for past $e^+ e^-$ colliders such as LEP or SLC.
 
The dominant channel is pair production via Drell-Yan and the vector boson fusion (VBF) pair production 
via gauge couplings is found to be subleading to Drell-Yan \cite{Agugliaro:2018vsu,Banerjee:2022xmu}.  
VBF single production is generated via topological anomalies, hence it is suppressed by a 
small anomaly coupling.
Drell-Yan single production could also be present if the pNGBs couple to quarks. However, it is expected
that pNGBs in models with partial compositeness couple only very weakly to light quark flavours 
as the couplings are roughly proportional to the quark mass. Consequently, the dominant couplings 
involve third generation quarks and the neutral pNGBs can be singly-produced via gluon fusion 
similar to the Higgs boson.
Moreover, both neutral and charged pNGBs can be singly-produced in association with either $tt$ or $tb$. 
This can provide a relevant contribution if the couplings are large enough.

We will present here recent results from recent investigations presented in 
refs.~\cite{Banerjee:2022xmu,Cacciapaglia:2022bax} to which we refer for further details.
We will first discuss simplified models and then focus on a specific model based on $\SU(5)/\SO(5)$
to present 
the interplay between various channels. 
This class of models can be realised in the context of four-dimensional models with a microscopic  
description \cite{Ferretti:2013kya,Ferretti:2014qta,Belyaev:2016ftv} and emerges as the 
minimal symmetry pattern from the condensate $\langle\psi\psi\rangle$ of two EW-charged fermions 
if $\psi$ is in a real irreducible representation  of the confining gauge group, e.g.~$Sp(4)$. 
We note for completeness, that the properties of the confining gauge dynamics, based on $\Sp(4)$, has been studied on the Lattice 
with promising results~\cite{Bennett:2017kga,Bennett:2019jzz,Bennett:2019cxd,Bennett:2022ftz}.
Complementary information on the mass spectrum and decay constants of the composite states can also 
be  obtained using holographic techniques \cite{Erdmenger:2020lvq,Erdmenger:2020flu,Erdmenger:2023hkl}.

\section{Bounds on Drell-Yan pair-produced scalars in simplified models} \label{sec:modelindependent}

We start with bounds in simplified  models where we focus  on pair production, via the dominant 
Drell-Yan channels. We use parts of a simplified model which has been introduced in refs.\
\cite{Banerjee:2022xmu,Cacciapaglia:2022bax} to which we refer for the underlying Lagrangian.
 We extend the SM by colourless scalar states $S^0,S^{0\prime}, S^\pm,S^{\pm\pm}$ that are physical mass 
 eigenstates labelled by their electric charge. We include the minimal set of states with charge up to 
two that have all the possible couplings to the EW gauge bosons. In case of the neutral we include 
two neutral states with opposite parity and assume that none of the BSM scalars obtains 
a non-zero vev.

\begin{figure}
    \centering
    \includegraphics{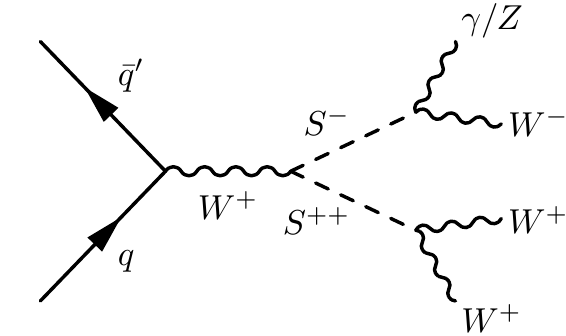} \qquad \includegraphics{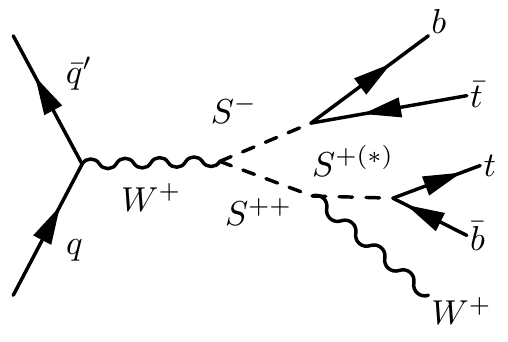}
    \caption{Examples of di-scalar channels from pair production via Drell-Yan processes with subsequent decays into SM particles.}
    \label{fig:pNGBpairgraphs}
\end{figure}

We investigate all combinations of scalar pairs produced at the LHC through the Drell-Yan processes:
\begin{equation}
\label{eq:DY}
pp \to  S^{\pm\pm} S^{\mp}\,,\, S^{\pm} S^{0(\prime)} \,,\,  
S^{++} S^{--} \,,\,S^{+} S^{-}
 \,,\,S^{0} S^{0\prime}\, .
\end{equation}
Together with the first tier decays of the scalar pairs into SM particles, 
these production processes yield many di-scalar channels, see for example \cref{fig:pNGBpairgraphs}. 
Charge-conjugated states belong to the same channel.
We consider two complementary scenarios for the decays of the scalars
\begin{enumerate}
\item The fermiophobic case, where 
couplings to SM fermions are absent at leading order and the dominant decays are into EW gauge bosons:
\begin{subequations}
	\begin{align}
		S^{++} &\to W^+ W^+\, , \\
		S^+ &\to W^+ \gamma, \, W^+Z \,,\\
		S^{0(\prime)} &\to W^+ W^-, \, \gamma \gamma, \, \gamma Z, \, ZZ.
	\end{align}
\end{subequations}
Combining the different Drell-Yan scalar pairs with the above decay channels leads to 24 
di-scalar channels, each containing four gauge bosons.
One sample process is shown in the left diagram of \cref{fig:pNGBpairgraphs}, while a complete list 
of all channels is shown in Table~\ref{tab:feriophobic}.
\item The fermiophilic case, where the scalars decay dominantly into a pair of third generation quarks:
\begin{subequations}
	\begin{align}
		S^{++} &\to W^+ t\bar b, \\
		S^+ &\to t\bar b, \\
		S^{0(\prime)} &\to t\bar t \,\,\,\mbox{ or }\,\, \, b\bar b.
	\end{align}
\end{subequations}
Note that doubly charged scalars cannot decay to two quarks due to their charge, but if they are part 
of an $\SU(2)_L$ multiplet, the three-body decay $S^{++}\rightarrow W^+S^{+*} \rightarrow W^+t\bar{b}$ is 
allowed. Consequently,  this yields 8 possible di-scalar channels for pair-produced scalars in this
scenario. One sample process is shown in the right diagram of \cref{fig:pNGBpairgraphs} and
a complete list  in \cref{tab:feriophilic}.

\end{enumerate}

\begin{table}[tb]
\centering
{\small \begin{tabular}{c|ccccc|}
fermiophobic & $\rescol{S^{++}}{S^{--}}$ & $\rescol{S^{\pm \pm}}{S^\mp}$ & $\rescol{S^+}{S^-}$ & $\rescol{S^\pm}{ S^{0(\prime)}}$ & $\rescol{S^0 }{S^{0\prime}}/\rescol{S^{0\prime}}{S^0}$ \\ \hline
$WWWW$ & $\rescol{W^+W^+}{W^-W^-}$ & - & - & - & $\rescol{W^+W^-}{W^+W^-}$ \\
$WWW\gamma$ & - & $\rescol{W^\pm W^\pm}{W^\mp \gamma}$ & - & $\rescol{W^\pm \gamma}{W^+ W^-}$ & - \\
$WWWZ$ & - & $\rescol{W^\pm W^\pm}{W^\mp Z}$ & - & $\rescol{W^\pm Z}{W^+ W^-}$ & - \\
$WW\gamma\gamma$ & - & - & $\rescol{W^+ \gamma}{W^- \gamma}$ & - & $\rescol{W^+W^-}{\gamma\gamma}$ \\
$WWZ\gamma$ & - & - & $\rescol{W^\pm \gamma}{ W^\mp Z}$ & - & $\rescol{W^+W^-}{\gamma Z}$  \\
$WWZZ$ & - & - & $\rescol{W^+ Z}{W^- Z}$ & - & $\rescol{W^+W^-}{ZZ}$ \\
$W\gamma\gamma\gamma$ & - & - & - & $\rescol{W^\pm \gamma}{\gamma \gamma}$ & - \\
$WZ\gamma\gamma$ & - & - & - & $\rescol{W^\pm}{} \{ Z \gamma\} \rescol{}{\gamma}$ & - \\
$WZZ\gamma$ & - & - & - & $\rescol{W^\pm}{} \{Z \gamma\} \rescol{}{Z}$ & - \\
$WZZZ$ & - & - & - & $\rescol{W^\pm Z}{ZZ}$ & - \\
$\gamma\gamma\gamma\gamma$ & - & - & - & - & $\rescol{\gamma\gamma}{\gamma\gamma}$\\
$Z\gamma\gamma\gamma$ & - & - & - & - & $\rescol{Z\gamma}{\gamma\gamma}$\\
$ZZ\gamma\gamma$ & - & - & - & - & $\rescol{Z}{}\{Z\gamma\}\rescol{}{\gamma}$\\
$ZZZ\gamma$ & - & - & - & - & $\rescol{ZZ}{Z\gamma}$ \\
$ZZZZ$ & - & - & - & - & $\rescol{ZZ}{ZZ}$ \\
 \end{tabular}}
 \caption{ \label{tab:feriophobic} Classification of the 24 di-scalar channels in terms of the 5 pair production cases (columns) and the 15 combinations of gauge bosons (rows) from decays. In the channels, the first two and second two bosons are resonantly produced. The notation $\{Z\gamma\} = \rescol{Z}{\gamma} + \rescol{\gamma}{Z}$ indicates the two permutations. Charge-conjugated states belong to the same di-scalar channel.}
\end{table}

\begin{table}[tb]
\centering
\begin{tabular}{c|ccccc|}
fermiophilic & $\rescol{S^{++}}{S^{--}}$ & $\rescol{S^{++}}{S^-}$ & $\rescol{S^+}{S^-}$ & $\rescol{S^+}{ S^{0(\prime)}}$ & $\rescol{S^0 }{S^{0\prime}}/\rescol{S^{0\prime}}{S^0}$ \\ \hline
$tttt$ & - & - & - & - & $\rescol{t\bar{t}}{t\bar{t}}$ \\
$tttb$ & - & - & - & $\rescol{t\bar{b}}{t\bar{t}}$ & - \\
$ttbb$ & - & - & $\rescol{t\bar{b}}{b\bar{t}}$ & - & $\rescol{t\bar{t}}{b\bar{b}}$ \\
$tbbb$ & - & - & - & $\rescol{t\bar{b}}{b\bar{b}}$ & - \\
$bbbb$ & - & - & - & - & $\rescol{b\bar{b}}{b\bar{b}}$ \\
$Wttbb$ & - & $\rescol{W^+t\bar{b}}{ b\bar{t}}$ & - & - & - \\
$WWttbb$ & $\rescol{W^+ t\bar{b}}{W^-b\bar{t}}$ & - & - & - & - \\
 \end{tabular}
 \caption{ \label{tab:feriophilic} Classification of the 8 di-scalar channels in terms of the 5 pair production cases (columns) and the 5 combinations of top and bottom from decays (rows). In cases with one or two doubly charged scalars, one always obtains $ttbb$ with one or two additional $W$'s, respectively. The charge-conjugated states are not shown.}
\end{table}

For the simulation of signal events we use the publicly available \texttt{eVLQ} model 
presented in  ref.~\cite{Banerjee:2022xmu}, which implements the simplified models as a \texttt{FeynRules} 
\cite{Alloul:2013bka} model at next-to-leading order in QCD. All events are generated at a 
centre-of-mass energy of $13$~TeV in $pp$ collisions.
For each di-scalar channel, we perform a scan over the scalar mass $m_S$; for channels involving 
two different scalars, we assume them to be mass degenerate. 
We generate $10^5$ events of Drell-Yan scalar pairs with decay into the target channel for each scan point.
We use \texttt{MadGraph5\_aMC@NLO} \cite{Alwall:2014hca} version 3.3.2 at NLO, in association with the parton densities in the \texttt{NNPDF 2.3} set \cite{Ball:2012cx,Buckley:2014ana}.
Afterwards, we interface the events with \texttt{Pythia8} \cite{Sjostrand:2014zea} for the decays of
the SM particles, showering and hadronisation. 
The resulting  signal events are then analysed with  \texttt{MadAnalysis5}  
\cite{Conte:2012fm,Conte:2014zja,Dumont:2014tja,Conte:2018vmg}  version 1.9.60 
and \texttt{CheckMATE} \cite{Drees:2013wra,Dercks:2016npn} version 2.0.34.  
Both tools reconstruct the events using \texttt{Delphes 3} \cite{deFavereau:2013fsa} and 
the anti-$k_T$ algorithm \cite{Cacciari:2008gp} implemented in \texttt{FastJet} \cite{Cacciari:2011ma}.
The exclusion associated with the events is calculated with the CL$_s$ prescription \cite{Read:2002hq}.
Moreover, we run the events against the SM measurements implemented in \texttt{Rivet} 
\cite{Bierlich:2019rhm} version 3.1.5 and extract exclusions from the respective 
\texttt{YODA} files using \texttt{Contur} \cite{Butterworth:2016sqg,Buckley:2021neu} version 2.2.1.
To present simplified model bounds, we determine the signal cross section $\sigma_{95}$ which is excluded at 95\% CL. We note for completeness, that the procedure differs somewhat between the tools and we
refer to ref.~\cite{Cacciapaglia:2022bax} for the details.
For each channel and each parameter point, we take the minimal value for $\sigma_{95}$ obtained 
from \texttt{MadAnalysis5}, \texttt{CheckMATE}, and \texttt{Contur} as the final bound. 
We do not attempt to combine them.

\begin{figure}[]
	\centering
	\begin{subfigure}{0.48\linewidth}
		\includegraphics[width=\linewidth]{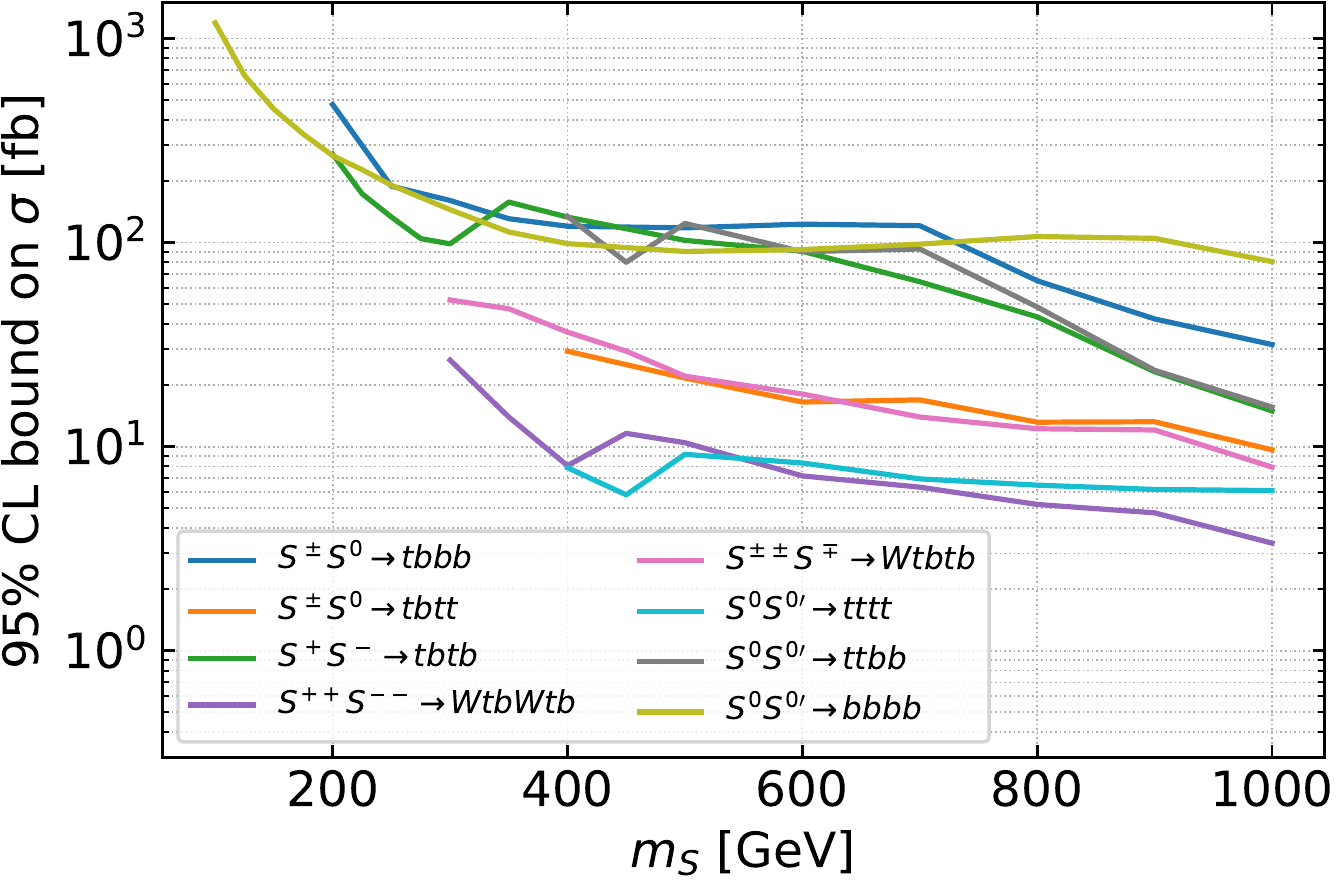}
		\caption{Scalar pair with decays to quarks}
		\label{fig:modelindependentquarks}
	\end{subfigure} \quad
	\begin{subfigure}{0.48\linewidth}
		\includegraphics[width=\linewidth]{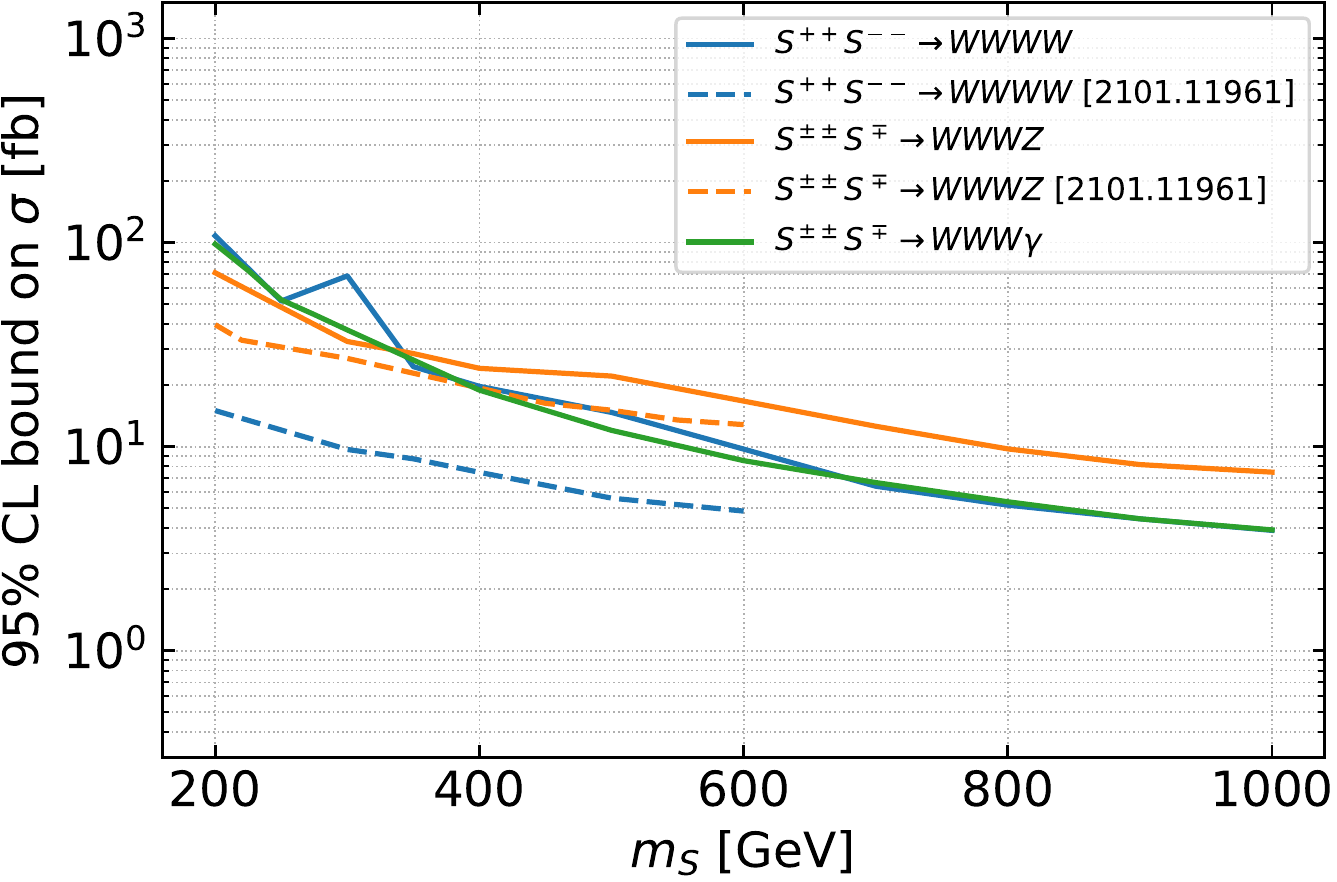}
		\caption{$S^{++} S^{--}$ and $S^{\pm\pm}S^\mp$ with di-boson decays}
		\label{fig:modelindependents12}
	\end{subfigure}  \vspace{2ex}

	\begin{subfigure}{0.48\linewidth}
		\includegraphics[width=\linewidth]{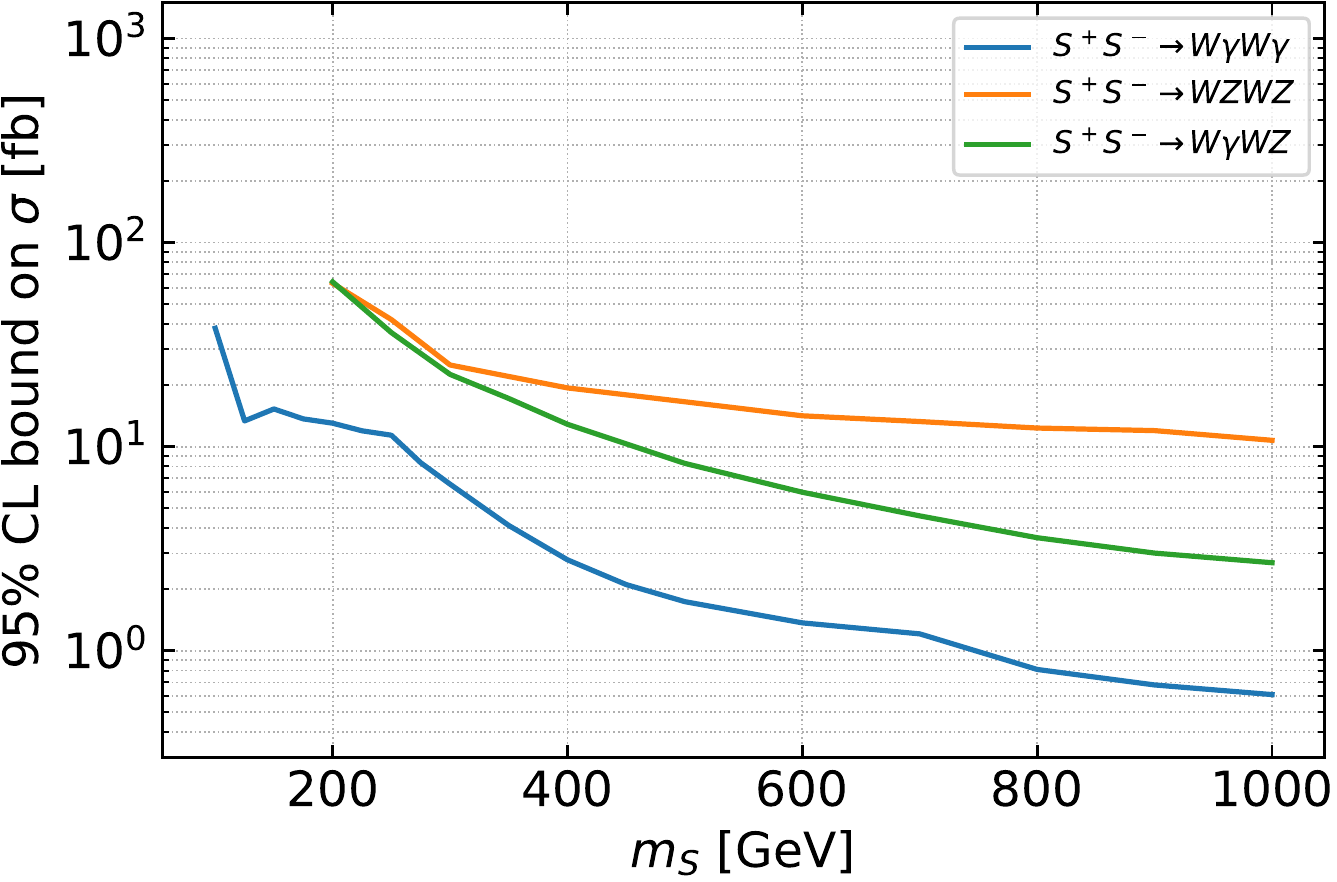}
		\caption{$S^{+} S^{-}$  with di-boson decays}
		\label{fig:modelindependents11s11}
	\end{subfigure} \quad
	\begin{subfigure}{0.48\linewidth}
		\includegraphics[width=\linewidth]{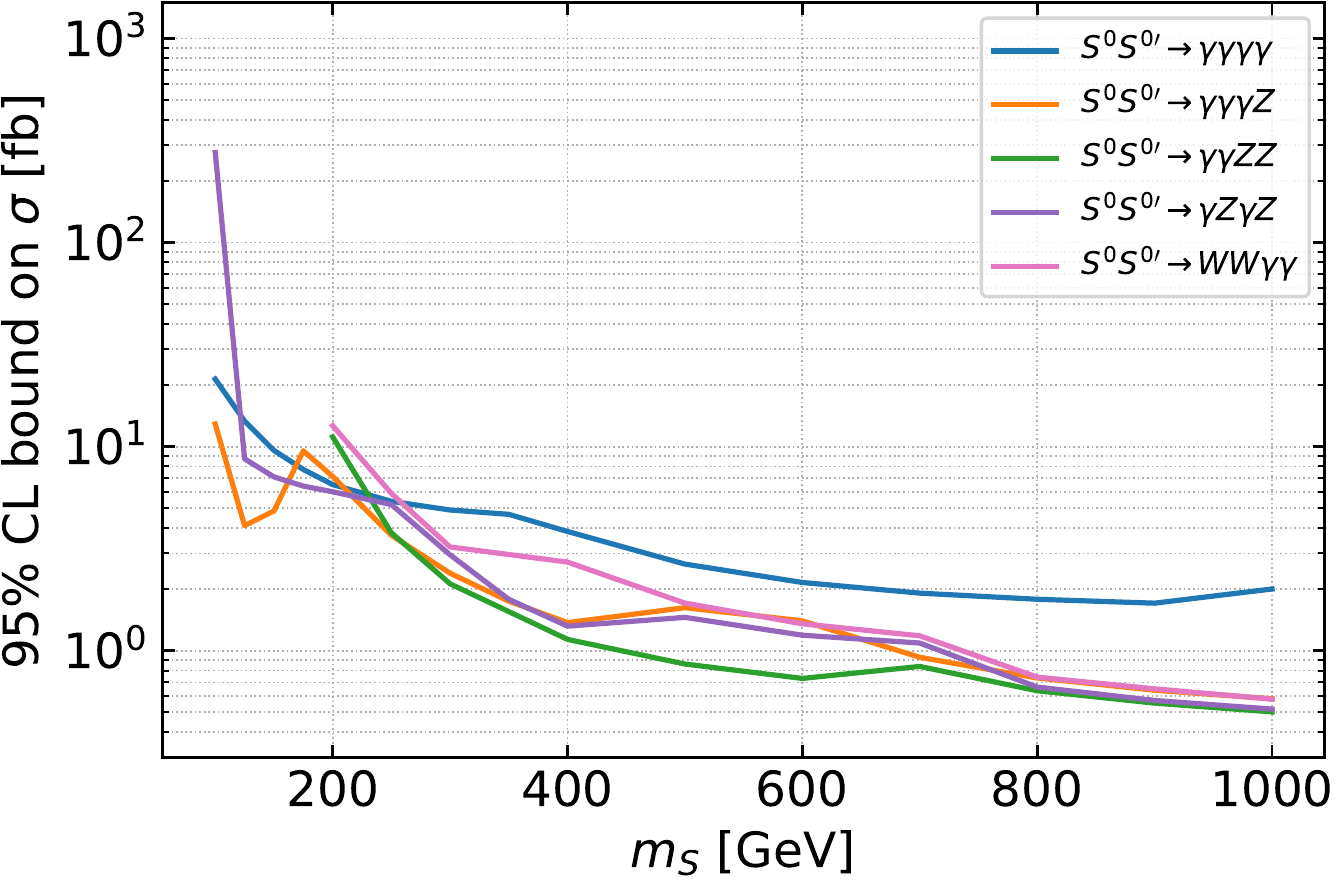}
		\caption{$S^{0} S^{0\prime}$  with di-boson decays with $\geq 2$ photons}
		\label{fig:modelindependents10s102aa}
	\end{subfigure} \quad
	\caption{Upper limits on the cross section of the di-scalar channels from Drell-Yan pair 
	production. The scalars decay to: (a) third generation quarks or (b)-(d) two vector bosons. Both
	 scalars are assumed to have the same mass. The analyses contributing to the bounds are
	 refs.~\cite{CMS:2019xud,ATLAS:2018nud,ATLAS:2021mbt,ATLAS:2021fbt,ATLAS:2021twp,ATLAS:2018zdn,ATLAS:2019gdh,CMS:2019xjf,CMS:2017abv,ATLAS:2020qlk,ATLAS:2021kog,ATLAS:2019gey,ATLAS:2018nci,CMS:2017moi}.} 
	\label{fig:modelindependent}
\end{figure}

We display results in \cref{fig:modelindependent}, where we present the simplified model bounds on 
the cross section for various di-scalar channels, i.e.\ bounds on the production cross section 
of the scalar pair times both branching ratios. 
We show in \cref{fig:modelindependentquarks} the bounds on the 8 di-scalar channels for the 
fermiophilic scenario, consisting of third generation quarks plus one additional $W$ boson 
per doubly charged scalar due to the 3-body decay of $S^{\pm\pm}$.
In channels with multiple top quarks, the dominant bounds stem from a search for $R$-parity 
violating supersymmetry \cite{ATLAS:2021fbt}, while various supersymmetric searches 
\cite{CMS:2019xjf,ATLAS:2018zdn,ATLAS:2021twp,ATLAS:2019gdh} and the generic search of 
ref.~\cite{CMS:2017abv} are relevant for the multi-bottom channels.

Figures \ref{fig:modelindependents12} to \ref{fig:modelindependents10s102aa} show bounds 
for various 
channels of the fermiophobic scenario which are split into three figures for the
sake of readability. 
In \cref{fig:modelindependents12} we display di-scalar channels with at least one doubly-charged scalar,
leading to at least 3 $W$ bosons plus a $W$, $Z$, or photon. The photon channel $WWW\gamma$ can be 
constrained using measurements of the $Z\gamma$ production cross section 
\cite{ATLAS:2019gey,ATLAS:2018nci}. The most relevant searches for the $WWWW$ and $WWWZ$ channels look 
for  multi-lepton final states \cite{CMS:2019xud, CMS:2017moi}. For these channels, the results of the 
ATLAS search \cite{ATLAS:2021jol} apply, and they are shown as blue and orange dashed lines. 
The bounds from this dedicated search are obviously stronger compared
the bounds obtained from our recasts of a large number of BSM searches which target different 
signatures and scenarios. This had to be expected and 
Figure~\ref{fig:modelindependents11s11} shows the di-scalar channels from $S^+S^-$ production. 
We note, that the bounds on $W\gamma W\gamma$ are by far the strongest, stemming from a search for 
gauge-mediated supersymmetry in final states containing photons and jets \cite{ATLAS:2018nud}. 
The main bounds for the channels $WZWZ$ and $W\gamma WZ$ stem from a multi-lepton search 
\cite{CMS:2017abv} and the $Z\gamma$ cross section measurements \cite{ATLAS:2019gey,ATLAS:2018nci},
respectively.
Finally, in \cref{fig:modelindependents10s102aa} we present the $S^0 S^{0\prime}$ channels 
containing at least 2 photons. The generic search \cite{ATLAS:2018zdn} and the
measurement of the $\gamma\gamma$-production cross section \cite{ATLAS:2021mbt}
constrain the $\gamma\gamma\gamma\gamma$ channel.
For the remaining channels, the most important analysis is a (multi-)photon search \cite{ATLAS:2018nud}.

\section{Bounds on the SU(5)/SO(5) pNGBs} \label{sec:su5so5}

Investigating simplified model is very useful approach as the limits can be applied to a broad class of models, at least to a certain extent. We investigate now a specific full model with an extended EW scalar sector, 
study the bounds on the full model and compare the results to estimates one can very quickly obtain 
by using the simplified model approach of the previous section.  We take the $\SU(5)/\SO(5)$ coset 
\cite{Agugliaro:2018vsu} as an example as it features a doubly charged scalar.
We first summarise some key elements and discuss some details of the underlying LHC phenomenology. 
For detailed discussions and the underlying couplings we refer to 
refs.~\cite{Banerjee:2022xmu,Cacciapaglia:2022bax}.
The pNGBs of the EW sector form a $\mathbf{14}$ of $\SO(5)$ which decomposes with respect to 
the custodial $\SU(2)_L \times \SU(2)_R$ as 
\begin{equation}
	\mathbf {14} \to (\mathbf 3, \mathbf 3) + (\mathbf 2, \mathbf 2) + (\mathbf 1, \mathbf 1) \,.
\end{equation}
We identify the $(\mathbf 2, \mathbf 2)$ with the Higgs doublet.
The bi-triplet can be decomposed under the custodial $\SU(2)_D \subset \SU(2)_L \times \SU(2)_R$ 
as \cite{Agugliaro:2018vsu}
\begin{equation}
	(\mathbf 3, \mathbf 3) \to \mathbf 1+\mathbf 3+\mathbf 5 \equiv \eta_1 + \eta_3 + \eta_5\,,
\end{equation}
with
\begin{equation} \label{eq:SU2Dbasis}
	\eta_1 = \eta_1^0, \quad \eta_3 = (\eta_3^+, \eta_3^0, \eta_3^-), \quad \, 
	\eta_5 = (\eta_5^{++}, \eta_5^+, \eta_5^0, \eta_5^-, \eta_5^{--}).
\end{equation}
This basis is suggested by the fact that the vacuum of the strong sector preserves the custodial 
$\SU(2)_D$. In the following we neglect a possible mixing and assume 
that the three multiplets have common masses $m_1$, $m_3$ and $m_5$, respectively, 
to simplify the analysis. Mass differences are due to the EW symmetry breaking, 
hence one naively expects a relative mass split of the order $v/m_i$ ($i=1,3,5$) with $v$ being the 
vev of the Higgs boson.

The LHC signatures of pNGB pair production depend strongly on whether the pNGBs are fermiophilic 
or fermiophobic as already mentioned above. We start with a brief discussion of the fermiophobic case
and refer to ref.~\cite{Cacciapaglia:2022bax} for further details.
The singly charged states decay as
\begin{equation}\label{eq:eta35pdec}
    \eta_{3,5}^+ \to W^+ \gamma,\, W^+ Z\,,
\end{equation}
with dominant photon channel as Br$(\eta_{3,5}^+\to W^+\gamma)\approx \cos^2 \theta_W \approx 78\%$ \cite{Agugliaro:2018vsu} for both multiplets in case of a small mass split between the multiplets.
The neutral singlet and quintuplet  decay dominantly as
\begin{equation}\label{eq:eta150dec}
    \eta_{1,5}^0 \to \gamma\gamma,\, \gamma Z,\, ZZ\,,
\end{equation}
with comparable branching ratios, again for small mass split.
The only available decay channel for the doubly charged pNGB in the quintuplet is
\begin{equation}\label{eq:eta5ppdec}
    \eta_5^{++} \to W^+ W^+.
\end{equation}
Finally, the $\eta_3^0$ is CP-even and thus undergoes three-body decays via off-shell pNGBs:
\begin{subequations}\label{eq:eta30dec}
\begin{alignat}{2}
    &\eta_3^0 \to W^+ W^- \gamma, \, W^+ W^- Z \quad &\text{ via } \eta_{3,5}^{\pm (*)}\,, \mbox{ and } \label{eq:eta30dec+}\\
    &\eta_3^0 \to Z \gamma \gamma, \, Z Z \gamma, \, ZZZ \quad &\text{ via } \eta_{1,5}^{0 (*)}\,. \label{eq:eta30dec0}
\end{alignat}
\end{subequations}
The discussion so far applies to the lightest multiplet and also covers scenarios 
where the multiplets are very close in mass. 
However, there could be a sizeable mass split. In such a case, cascade decays 
from one multiplet into a lighter one and a (potentially off-shell) vector boson become important. 
Taking the case $m_5>m_3>m_1$ as an example, we have
\begin{subequations}\label{eq:etachaindec}
\begin{gather}
    \eta_5^{++} \to W^{+(*)} \eta_3^+, \qquad
    \eta_5^+ \to Z^{(*)} \eta_3^+,\, W^{+(*)} \eta_3^0, \qquad
    \eta_5^0 \to W^{\pm(*)} \eta_3^\mp, \, Z^{(*)} \eta_3^0, \\
    \eta_3^+ \to W^{+(*)} \eta_1^0, \qquad \eta_3^0 \to Z^{(*)} \eta_1^0.
\end{gather}
\end{subequations}
One finds that both classes of decays are of similar importance once the mass split is between 
$30$ and $50$~GeV \cite{Cacciapaglia:2022bax}. We note for completeness, that the $\eta_5$ multiplet
does not couple to $\eta_1^0$ in the model considered.

One expects in the fermiophilic case that the couplings  scale like the quark masses, e.g.
\begin{align}\label{eq:couplingscaling}
\kappa^{\eta^0_i}_{t} = c_t^i\,
\frac{m_t}{f} \quad , \quad
\kappa^{\eta^0_i}_{b} = c_b^i\,
\frac{m_b}{f} \quad \text{and} \quad
\kappa^{\eta^+_j}_{tb} = c_{tb}^j\,
\frac{m_t}{f}\,,
\end{align}
where $f$ is the decay constant of the $\mathbf{14}$-plet and the $c$ coefficients are of order one.
In this case the decays to third generation quarks dominate over the loop-level anomaly-induced 
decays into two vector bosons or the three-body decays discussed above and, thus,
 we  consider  the decays
\begin{align}
    \eta_{3,5}^+ \to t \bar b, \qquad \eta_{1,3,5}^0 \to t \bar t,\, b\bar b\,.
\end{align}
From \cref{eq:couplingscaling}, the $t\bar t$ channel dominates over $b\bar b$ above threshold.
In the case of $\eta^{++}_5$, the three-body decay
\begin{equation}
    \eta_5^{++} \to W^+ t \bar b
\end{equation}
via an off-shell $\eta_{3,5}^+$ dominantes over the decay to $W^+ W^+$. 
In case of $m_5 > m_3$ also the decay $\eta^{++}_5 \to W^{+(*)} \eta^+_3$ becomes important 
\cite{Cacciapaglia:2022bax}.

\begin{figure}
    \centering
    \begin{subfigure}{0.48\linewidth}
        \includegraphics[width=\linewidth]{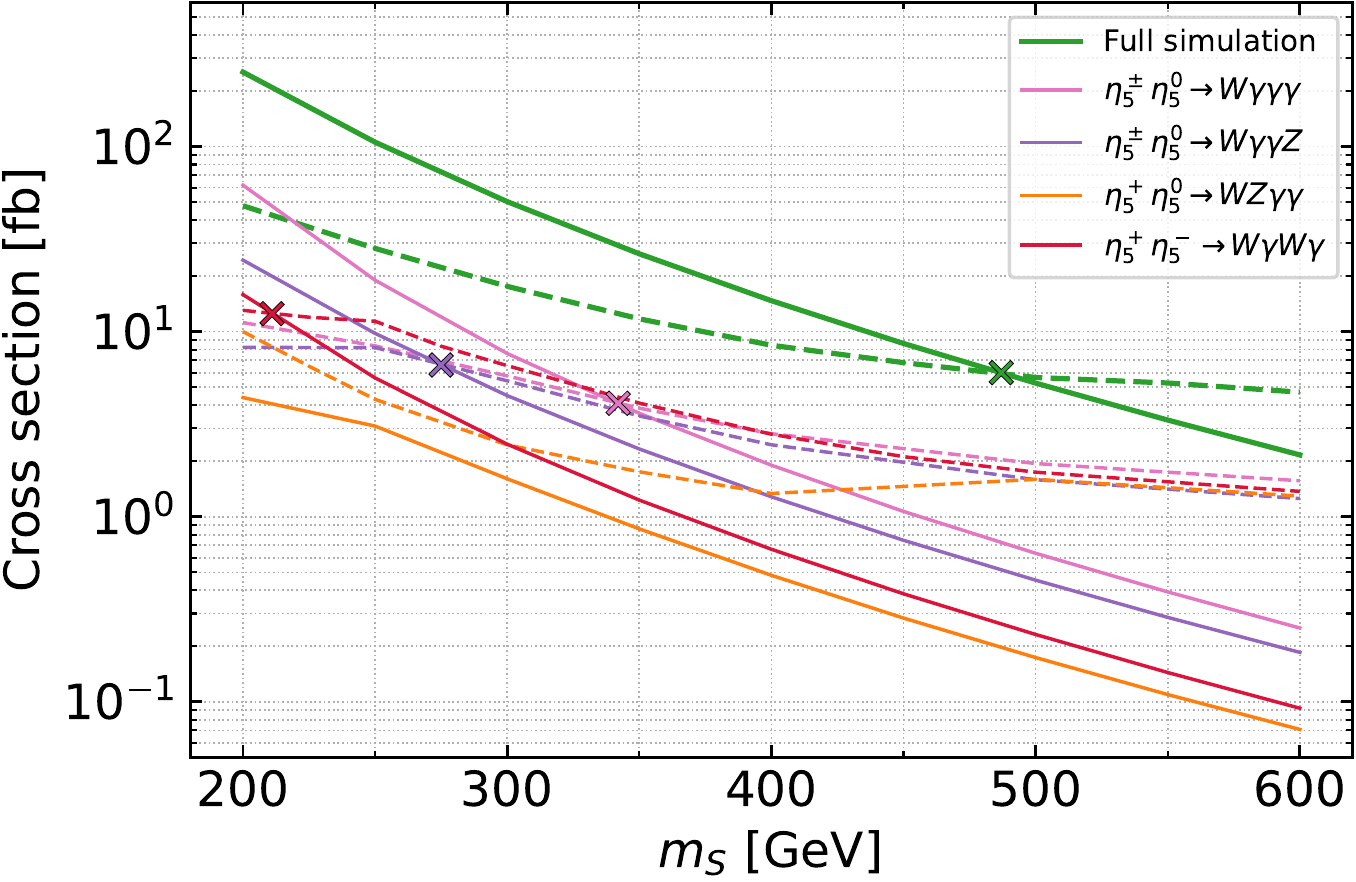}
        \caption{Bounds from individual channels}
        \label{fig:illustrationindividual}
    \end{subfigure} \quad
    \begin{subfigure}{0.48\linewidth}
        \includegraphics[width=\linewidth]{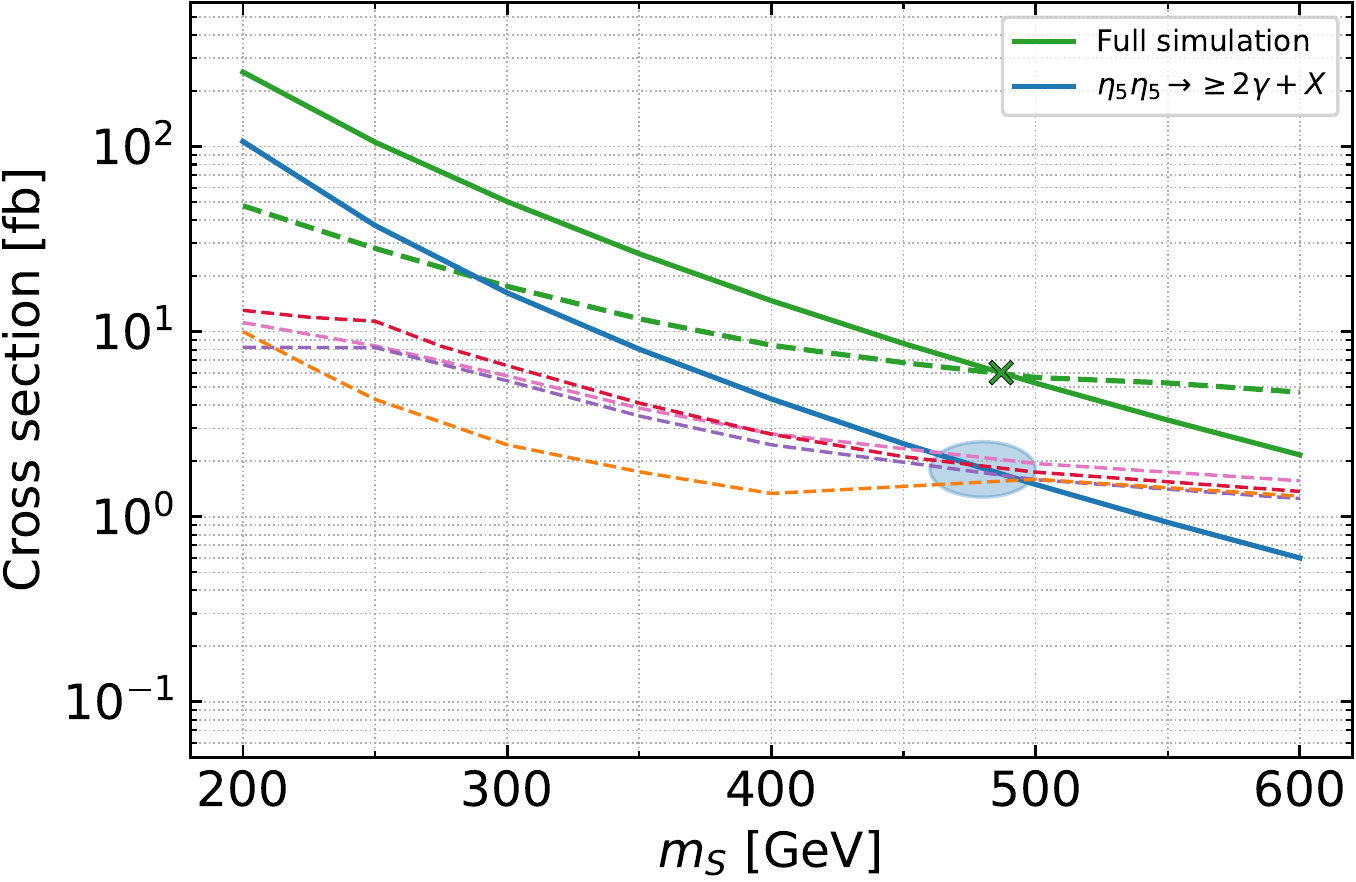}
        \caption{Bounds from sum of multiphoton channels}
        \label{fig:illustrationcombined}
    \end{subfigure} \vspace{2ex}
    
    \caption{Application of the model-independent bounds to a specific model, the custodial quintuplet $\eta_5$ from the $\SU(5)/\SO(5)$ coset.
    In (a) we determine the bounds from the dominant individual channels by comparing the cross section time branching ratio from the model (solid) with the upper limits from \cref{fig:modelindependent} (dashed).
    In green we show the results of a full simulation. The blue line in (b) is the sum of the individual multi-photon cross sections shown in (a).
    Further details are given in the text.}
    \label{fig:illustration}
\end{figure}

We consider in a first step only the quintuplet $\eta_5$ and apply the simplified model bounds 
from the previous section. In \cref{fig:illustrationindividual} we compare the cross section 
times branching ratio of all multi-photon final states (solid lines) with the corresponding 
bounds from \cref{fig:modelindependent} (dashed lines).
From the individual channels one finds that masses below $340$~GeV are excluded, with the strongest bound 
coming from the channel $\eta_5^\pm \eta_5^0 \to W\gamma\gamma\gamma$. We perform in addition a full
simulation in which all states contained in the quintuplet are pair-produced and decayed.  
The solid green line denotes the sum over all pair production cross sections of the quintuplet and
the dashed green line the corresponding bound, i.e.\ the sum of scalar pair production cross sections 
that would be needed in order to exclude the convolution of all decay channels from quintuplet states. 
As can be seen, one obtains a bound of $485$~GeV on the mass $m_S$ which is significantly higher than 
the bounds obtained from individual channels. 
The apparent discrepancy between simplified models and the full simulation can be understood from the 
fact that all   multi-photon channels populate the same signal region of the search presented in ref.~\cite{ATLAS:2018nud}. 
Adding up the various signal cross sections with two or more photons gives the blue line in 
\cref{fig:illustrationcombined}.
This summed cross section  yields an estimated bound on $m_S$ of $460-500$~GeV when compared
 with the bounds from different multi-photon channels (shaded area in \cref{fig:illustrationcombined}).
This is in agreement with the result of the full simulation. This example shows the usefulness 
and limitations of the simplified model bounds and demonstrate how they can be combined 
in the context of a particular model.

In a second step, we take all multiplets into account and consider two scenarios which
are characterised by varying a single mass scale $m_S$:
\begin{subequations}\label{eq:scenarios}
\begin{alignat}{4}
    \text{S-eq:}\quad &m_3 = m_S -2~\mathrm{GeV},\quad &&m_5 = m_S,\quad &&m_1 = m_S+2~\mathrm{GeV}\,; \\
    \text{S-135:}\quad &m_1 = m_S - 50~\mathrm{GeV},\quad &&m_3 = m_S,\quad &&m_5 = m_S + 50~\mathrm{GeV}\,.
\end{alignat}
\end{subequations}
The choice of $50$~GeV is motivated by the fact that the mass splits are expected to be a 
fraction of the Higgs vev.
The phenomenology differs in the two cases:
In scenario S-eq, all particles decay via the anomaly and $\eta_3^0$ exhibits three-body decays.
We introduce a small mass split of 2~GeV to avoid a cancellation for some $\eta_3^0$ decays as discussed
in ref.~\cite{Cacciapaglia:2022bax}. 
In scenario S-135  the heavier states decay into the next lighter states or di-bosons, while the lightest state only has anomaly induced decays.
\begin{figure}
    \centering
    \begin{subfigure}{0.48\linewidth}
        \includegraphics[width=\linewidth]{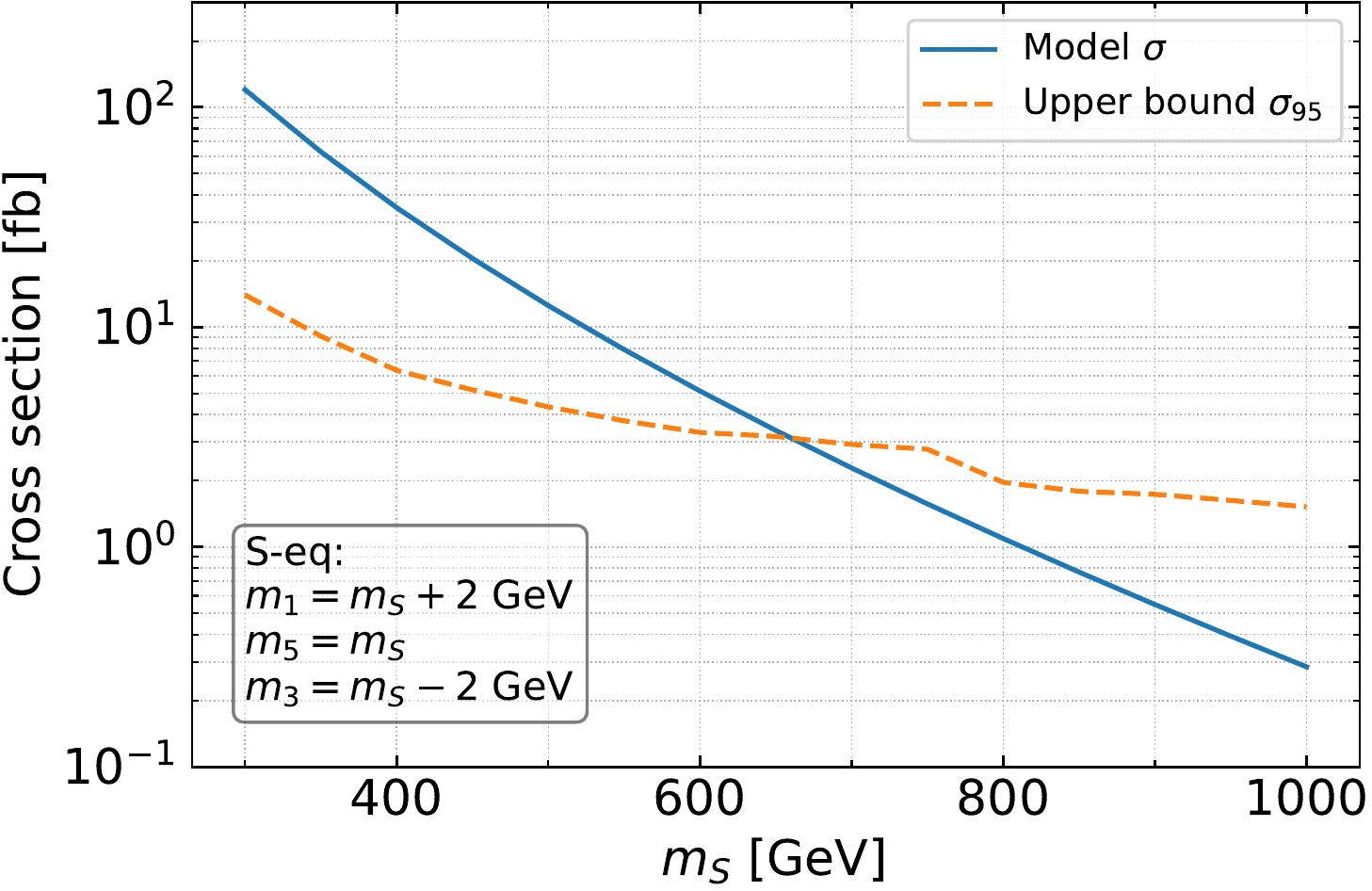}
        \caption{Scenario S-eq}
    \end{subfigure} \quad
    \begin{subfigure}{0.48\linewidth}
        \includegraphics[width=\linewidth]{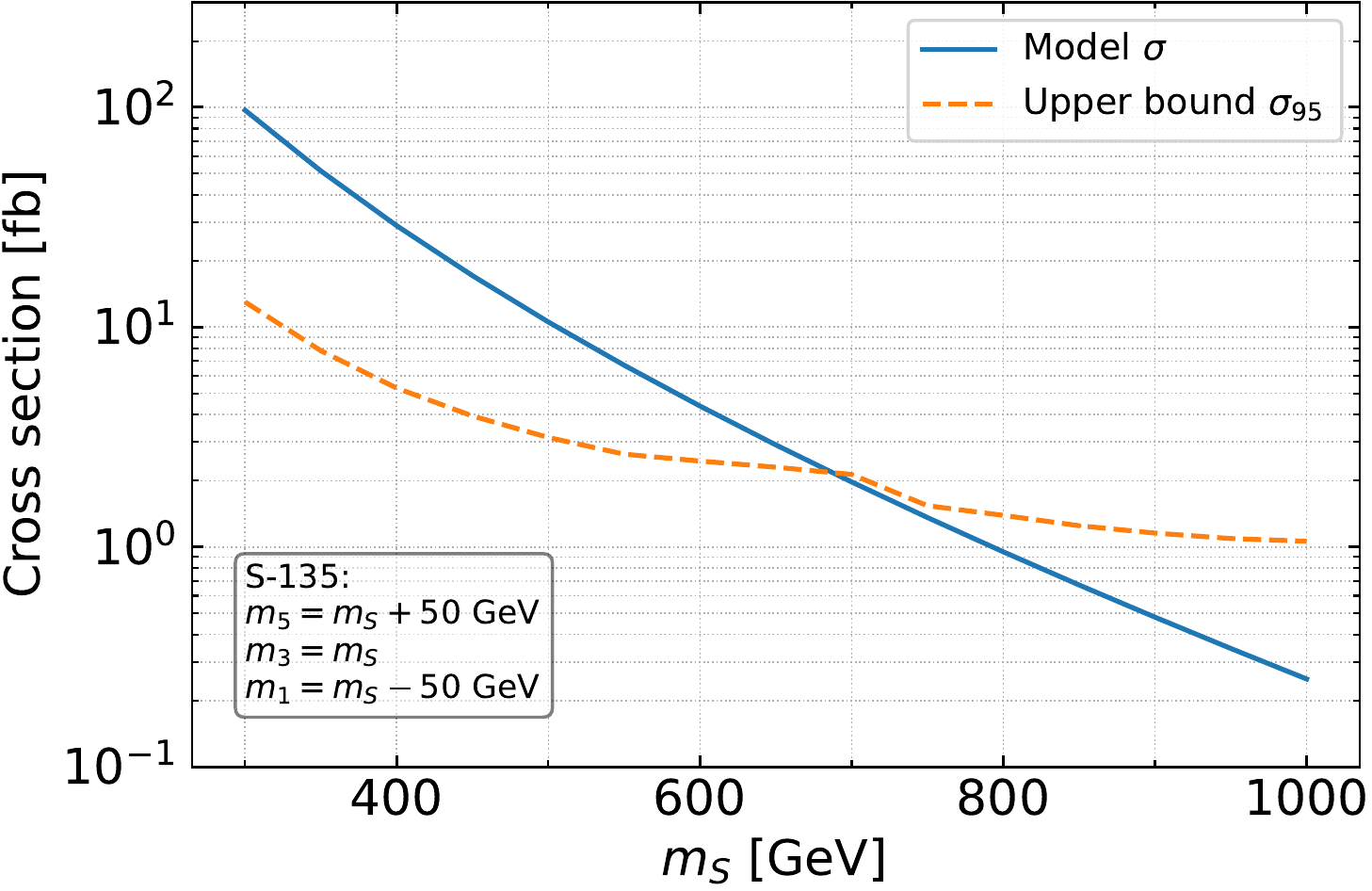}
        \caption{Scenario S-135}
    \end{subfigure} 
    \caption{Bounds on the pNGB masses for the Drell-Yan production of the full bi-triplet for 
    the benchmark mass spectra defined in \cref{eq:scenarios}. In (a) all masses are 
    approximately equal and in (b) there is a 50~GeV mass split between the multiplets.}
    \label{fig:1dimbounds-wzw}
\end{figure}
We present the bounds on the mass parameter $m_S$ for the two scenarios in \cref{fig:1dimbounds-wzw}.
In orange the sums over all scalar pair production cross sections $\sigma_{95}$ is shown that would 
be needed to exclude the model at 95\% CL at each parameter point.
The strongest bounds come from multi-photon channels, with ref.~\cite{ATLAS:2018nud} being the 
dominant analysis. Note, that the kink in $\sigma_{95}$ is due to a change in dominant signal 
region within the same analysis. In blue, the actual sum over all pair production cross sections 
in this model is drawn.
The different bounds for these scenarios considered are due to the relative size of the cross section for the triplet and quintuplet.

We turn now to the scenarios in which the pNGBs couple dominantly to quarks.
In these scenarios, one has single scalar production
via the processes
\begin{align}
    pp\to S^0 t\bar{t} \quad
    \text{ and } \quad 
    pp\to S^\pm t b \,
\end{align}
induced by strong interactions. In addition, the couplings of the neutral scalars to quarks 
induce couplings to gluons and photons at the one-loop level leading to processes like
\begin{align}
    pp \to S^0  \to t\bar{t} \quad \text{ and } \quad
      pp \to S^0  \to \gamma \gamma \,.
\end{align}
It turns out that presently available searches can constrain these channels for masses of up to 500 GeV
only if the $c$-factors in \cref{eq:couplingscaling} are close to $5$ and a decay constant $f$ of 1~TeV
which is only a very small part of the available parameter space.

\begin{figure}
    \centering
    \includegraphics[width=0.6\linewidth]{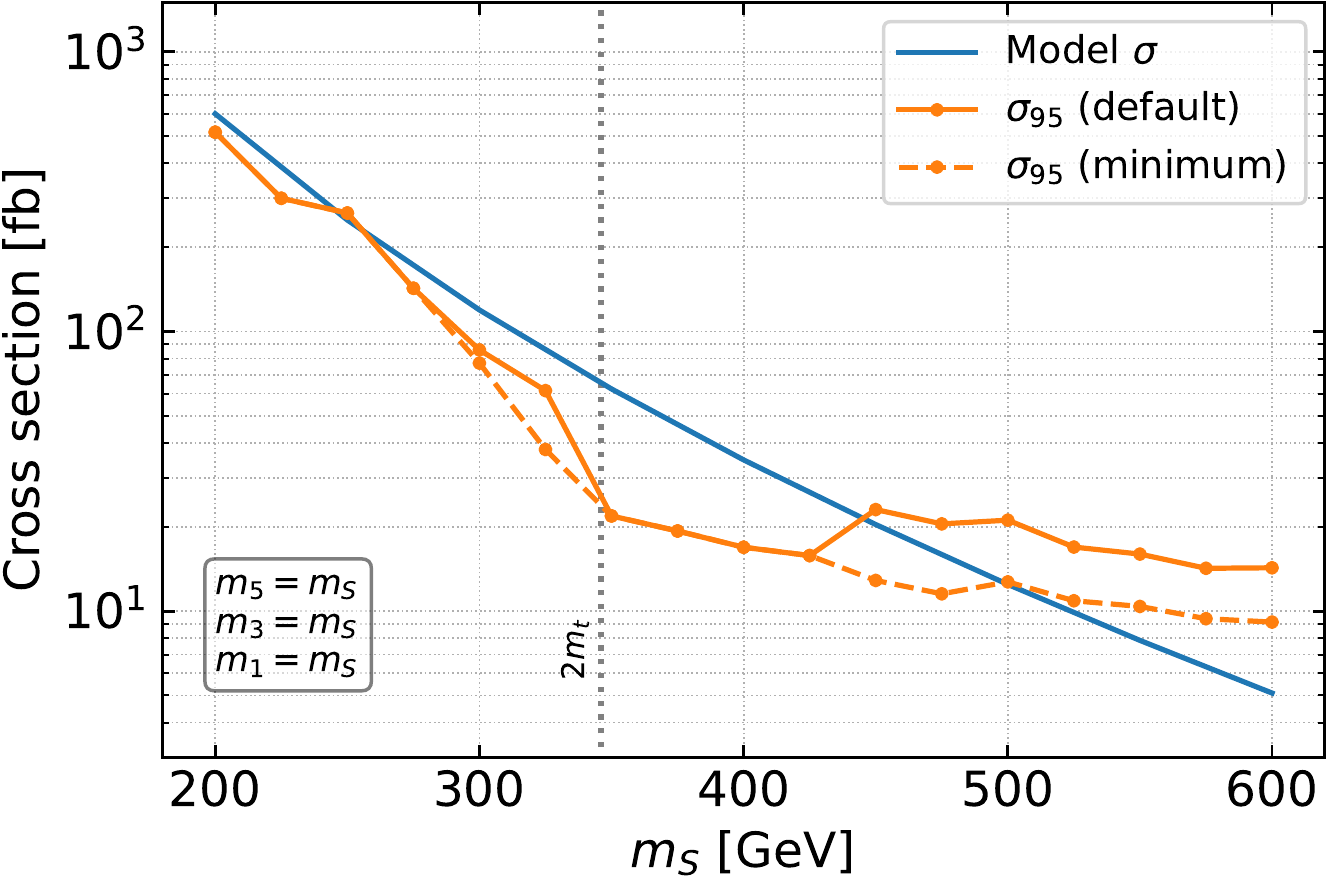}    
    \caption{Bounds on the pNGB masses for the Drell-Yan production of the full bitriplet with decays to third-generation quarks.}
    \label{fig:1dimbounds-quarks}
\end{figure}

We now turn to Drell-Yan pair production and display our results in \cref{fig:1dimbounds-quarks}. 
We have assumed that all pNGBs are mass degenerate and that all $c$-factors are 1. Note, that neither 
branching ratios nor production cross sections depend on $f$. 
The blue line gives the total cross section summing over all pNGBs irrespective of their decay modes. 
The orange lines give the exclusion when considering all possible channels. The exclusion is dominated 
by the results of ref.~\cite{ATLAS:2021fbt} implemented in \texttt{CheckMATE}. 
\texttt{CheckMATE} uses the signal region with the strongest expected bound and reports the 
corresponding observed bound as final result. One obtains the bound given by the solid orange line
using this standard procedure. This can lead to difficulties in cases for which observed and 
expected bounds differ significantly. This is the reason for the kinks at $m_S=350$~GeV and $450$~GeV. 
However, modifying the procedure such that always the strongest observed bound is taken, 
one obtains a smoother curve shown by the dashed orange line, see \cite{
Cacciapaglia:2022bax} for further details. 

We note for completeness, that models with and gauge/fermionic underlying dynamics 
\cite{Ferretti:2013kya,Ferretti:2014qta,Ferretti:2016upr} also predict pNGBs which are strongly
interacting. The $\SU(5)/\SO(5)$ coset is for example realized in the model M5 of ref.~\cite{Belyaev:2016ftv}
which predicts beside the electroweak pNGBs discussed above also a color triplet pNGB $\pi_3$ and
a color octet pNGB $\pi_8$. The $\pi_8$ decays either dominantly into $tt$ or $gg$ depending
on whether we are in a fermiophilic or a fermiophobic scenarios, respectively. In both cases
currently available analyses give a bound of about 1.1~TeV on its mass \cite{Cacciapaglia:2015eqa}.
The $\pi_3$ has the same quantum numbers as a right-handed stop, the supersymmetric partner of the right-handed top-quark, 
and its phenomenology depends on the mass spectrum of the baryonic bound states 
\cite{Cacciapaglia:2021uqh}. This baryonic bound states contain two color neutral fermions denoted
by $\tilde b$ and $\tilde h$ in \cite{Cacciapaglia:2021uqh} where the first one is a gauge singlet
and the second an $SU(2)_L$ doublet with hypercharge 1/2. It is possible that this states, being
color singlets, can be lighter than $\pi_3$. In such a scenario the $\pi_3$ decays into one of these fermions and an
SM-fermion:
\begin{align} \label{eq:pi3decayA}
\pi_3 \to t  \tilde B,\ t \tilde{h}^0,\ b \tilde{h}^+\,.
\end{align}
This resembles the decays of stops in supersymmetric models.
In principle, decays into lighter families, like $c \, \tilde B$ and $u \, \tilde B$, are also 
possible, but in the spirit of composite Higgs models we expect those to be strongly suppressed.
Consequently, LHC bounds from stop  searches can be directly applied 
\cite{CMS:2019zmd,ATLAS:2020xzu,CMS:2021eha}. For large mass differences this gives a bound of about 1.3 TeV. 
We note fore completeness, that in case of small mass difference between $\pi_3$ and $\tilde B$
three-body decays via an off-shell top-quark would become important similar to supersymmetric models 
\cite{Porod:1996at,Porod:1998yp,Boehm:1999tr}. 
In case that $\pi_3$ is lighter than these baryons, lepton or baryon number violating interactions need to be included in order to avoid a stable $\pi_3$ which require an extension of this model 
\cite{Cacciapaglia:2021uqh}. 
The simplest possibilities are 
\begin{align} \label{eq:pi3decayBb}
\pi_3 \to \bar{d}_i \, \bar{d}_j \quad \text{ with } d_i = d,s,b \text{ and } i\ne j 
\end{align}  
or
\begin{align} \label{eq:pi3decayBl}
\pi_3 \to u_i \, \nu_{l_j} \,,\, d_i \, l_j \quad \text{ with } u_i = u,c,t \text{ and } 
l_j = e,\mu,\tau\,. 
\end{align}  
The former violates baryon number whereas the latter violates lepton number.
Note, that only one of the two interaction types can be present as otherwise the  proton could
decay at a rate incompatible with experiment. In case of lepton number violation, one can use searches
for leptoquarks and finds a bound of 1.4-1.5~TeV depending on whether final states with a $\tau$ or
$e,\mu$ dominates \cite{ATLAS:2022wcu,ATLAS:2023uox}.

\section{Conclusions}

In this contribution we have presented bounds on the Drell-Yan pair production of 
scalar bosons that carry electroweak charges at the LHC. We first consider various
channels in a simplified model approach. These channels contain either four 
vector bosons or  top/bottom quarks in the final states. These two scenarios arise 
from fermiophobic and fermiophilic models, respectively. The limits, 
presented in \cref{fig:modelindependent}, can be applied to any model with an 
extended Higgs sector dominated by pair production.

In addition we have taken as a concrete example a composite Higgs model based on the 
coset $\SU(5)/\SO(5)$, which features a custodial bi-triplet. 
We show that while the limits on individual channels lead to relatively weak bounds on the 
scalar masses, significantly stronger bounds can be obtained by combining various 
pair production channels. Considering various benchmark scenarios, limits on 
the scalar mass scale around $500-700$~GeV have been established in the fermiophobic case.
The bounds are close to $500$~GeV in scenarios in which decays into top 
and bottom quarks dominate.

\section*{Acknowlegments}

I thank the organizers of this meeting for an inspiring atmosphere. I also thank 
J.~Butterworth, G.~Cacciapaglia, T.~Flacke, M.~Kunkel, R.~Str\"ohmer and L.~Schwarze for discussions.

\end{document}